\documentstyle[12pt,epsf]{article}
\begin{document}
\begin{titlepage}

\begin{flushright}
{\bf MC-TH-95/18}\\
{December 1995}
\end{flushright}

\vspace{1cm}

\begin{center}
\begin{large}
{\bf Data Unfolding in $W$ Mass Measurements at LEP2 }

\vspace{1cm}

{V.Kartvelishvili$~{}^{a)}$}\\
\bigskip
\end{large}
{\em Department of Physics and Astronomy,\\
Schuster Laboratory, University of Manchester,\\
Manchester M13 9PL, U.K.\\}

\vspace{0.5cm}

\begin{large}
{R.Kvatadze$~{}^{a)}$}\\
\bigskip
\end{large}
{\em Joint Institute for Nuclear Research,\\
Dubna, Moscow Region, RU-141980, Russia}\\

\vspace{2cm}

{\bf Abstract}
\end{center}

\vspace{0.5cm}

\begin{narrower}
\noindent
The use of an unfolding procedure is proposed as an alternative method
of extracting the $W$ boson mass from the data measured at LEP2,
which may improve the accuracy of
this measurement. The benefits of the direct
unfolding method based on the Singular Value Decomposition of the response
matrix are demonstrated on the example of $W$ mass determination
from the charged lepton energy spectra.
\end{narrower}

\vspace{4cm}

\noindent{---------------------}\\
${}^{a)}$ On leave from High Energy Physics Institute, Tbilisi State
University, Tbilisi, GE-380086, Republic of Georgia.

\end{titlepage}

\newpage
\section{Introduction}
\vspace{0.5cm}

One of the main physics motivations for the LEP
 energy upgrade is the precise measurement of the $W$-boson mass.
 This is necessary to test the Standard
 Model of the electroweak interactions, and together with the
 top quark mass would allow to set more significant limits for 
 the Higgs boson mass.
 The existing information on the $W$-boson mass comes from the
 measurements in $\bar p p$ interactions
 at the Tevatron {\cite{tn1}} and
 %D0 at FNAL, as well as  UA2 results at 
 CERN {\cite{tn2}}.
 In these experiments, transverse momentum distributions of
 the charged leptons, missing transverse momentum spectra, and
 transverse mass distributions of the charged lepton-neutrino
 system were used to obtain the $W$ mass, resulting in the world
 average $80.26\pm 0.16$ GeV {\cite{waw}}. 

Various methods have been proposed for the precision
 determination of the $W$ mass in $e^+e^-$ annihilation at
 LEP2 {\cite{tn3,tn4}}. Extensive tests and studies resulted in the
 following conclusions:

\begin{itemize}

\item {The direct reconstruction of the $W$ mass from the final state
 particles is considered to be the most promising method,
 if one uses the four constraints from the energy-momentum
 conservation and the assumption
 that the $W$'s in the intermediate state have equal masses.
 Two decay channels --- $q \bar q q \bar q$ (four jets) and
 $q \bar q l \nu$ (two jets plus leptons) --- can be used in this
 analysis.
 If colour reconnection {\cite{tn5}} and Bose-Einstein correlations
 {\cite{tn6}}
 do not lead to additional significant errors, then the
 $W$ mass can be measured with the precision of about 60 MeV
 (statistical) and 40 MeV (systematic), for a single
 LEP2 experiment {\cite{tn4}}. Otherwise, one should concentrate on
 the decay channel of two jets plus leptons, which is
 free of these complications, but results in a larger statistical error
 of $\approx 80$ MeV per experiment.}

\item {$W$ mass measurement from
 the rise of the $W^+ W^-$ cross section near the threshold gives
 the error competitive, but not better than the direct
 reconstruction method.}

\item {Measurement of the charged lepton energy
 spectra does not allow one to achieve comparable precision
 for $W$ mass at any energy of LEP2.}

\end{itemize}

 In this paper we consider the semileptonic decay channel $jet+jet+l+\nu$. 
 The actual measurement gives the following information:

\begin{itemize}
\item{The distribution of the hadronic and electromagnetic energy
 deposited into the calorimeters.}
\item{Momenta of charged hadrons.}
\item{The energy-momentum of the charged lepton.}
\end{itemize}

Obviously, the complete event is lost if the charged
 lepton escapes detection, hence the geometric coverage is vital.
 The major error comes from the limited acceptance of hadron
 detection and finite resolution of energy-momentum
 measurements, so that the invariant mass
 of the measured hadronic system is significantly lower than the
 mass of the decayed
 $W$-boson. This also prevents  the overall energy-momentum
 conservation constraints to be used directly
 to determine the 4-momentum of the neutrino, unless the rescaling
 of the hadronic jet energies is performed.
 The overall kinematical fit of
 the whole event allows one to achieve the required accuracy only
 if the masses of the
 two $W$-bosons in the event are assumed to be equal to each other.
 So, each event in this approach gives just a single entry into the
 $W$ mass distribution instead of two, thus effectively reducing
 the available statistics.
 It is argued, however, that the gain in the precision of the resulting
 $W$-boson mass should justify the use of this procedure
 {\cite{tn3,tn4}}.

We suggest a different way of extracting the mass
 distributions of W-bosons from the same data. Our method is based
 on the direct unfolding of the measured data straight into
 the $W$ mass.
 This analysis requires a high statistics
 Monte Carlo simulation of the response matrix  describing the whole
 measurement process. The method does not use
 the assumption of equal $W$ masses and does not require any rescaling of
 measured hadronic jet energies. All necessary unfolding (i.e. corrections
 for efficiencies, acceptances and resolutions of various detectors used
 for the measurement) is performed in a consistent way in one go, in the
 framework of the regularized unfolding procedure based on the Singular
 Value Decomposition of the response matrix {\cite{svd}}.
 The unfolded $W$-boson mass distribution
 will have its natural decay width, not broadened by the detector
 resolution effects, and can be fitted to obtain $M_W$.

The suggested procedure is presented in some detail in the following
 section. The general description of the unfolding method used in this paper
 is given elsewhere {\cite{svd}}. The latter has been tested on various
 examples and is stable and reliable, provided the response matrix
 is known precisely enough. In order to apply the suggested procedure to
 the channel $jet+jet+l+\nu$ one has to have a detailed
 Monte Carlo simulation of this process, including full detector response
 for hadrons and the charged lepton,
 with the statistics of at least an order of magnitude
 larger than the expected number of measured events.
 In the absence of this information, in
 Section \ref{exa}
 we have restricted ourselves to the case when only
 the single charged lepton energy spectrum is used
 to unfold the $W$ mass distribution.
 This example is easier to simulate
 but is more difficult to unfold,
 as the $W$ mass distribution in this case is hidden behind a non-trivial
 functional dependence in addition to the usual detector effects.
 Nevertheless, it incorporates many features of the full
 problem, and demonstrates the
 ability of our method to deal with multi-dimensional distributions.
 Some conclusions are drawn out in Section \ref{dis}.

\section{Description of the procedure}

Consider the Monte Carlo simulation of the process
\begin{equation}
 e^+ + e^- \to W^+ + W^- \to jet + jet + l + \nu.
\end{equation}
 Let $M_1$ be the mass of the $W$ which decayed leptonically,
 while $M_2$  is the mass of the second $W$, which decayed hadronically.
 The distributions of both
 $M_1$ and $M_2$ essentially follow the Breit-Wigner curve,
 apart from phase space factors.
 Each generated event corresponds to an entry into a
 two-dimensional histogram $M_1~ vs.~ M_2$, and
 there is no reason why the  two projections of the latter 
 onto the axes $M1$ and $M2$
 should be different from each other, apart from obvious statistical
 fluctuations.

After the
 measurement of each event one gets the distribution of the measured
 energy over the calorimeter cells, and the momenta
 of charged hadrons and the charged lepton. 
 Define the initial four-vector $P\equiv(\sqrt{s},\vec{0})$, 
 the measured four-vector of the lepton
 $p_l\equiv(E_l,\vec{p}_l)$ and the
 summary four-vector of the whole measured hadronic system
 $p_h\equiv(E_h,\vec{p}_h)$.
 The initial energy $\sqrt{s}$ and the lepton mass $\sqrt{E_l^2-\vec{p}_l^2}$
 are known, so one has no more than 7 independent
 measured components, some of which
 do not contain any significant information on the masses of 
 the decaying $W$s. Four non-trivial invariant combinations of the 3
 vectors defined above can be combined to calculate the following quantities
 (lepton masses have been neglected):
\begin{itemize}
\item
 the invariant mass of the measured hadronic system $M_h$,
\begin{equation}
M_h^2=p_h^2;
\end{equation}
\item
the invariant mass of the system recoiling from the measured
hadrons $M_A$,
\begin{equation}
M_A^2\equiv (P-p_h)^2 = M_h^2 + \sqrt{s}(\sqrt{s}-2E_h);
\end{equation}
\item
the invariant mass of missing particles $M_{\mathrm{miss}}$,
\begin{equation}
M_{\mathrm{miss}}^2\equiv (P-p_h-p_l)^2=
M_A^2 - 2E_l(\sqrt{s}-E_h+|\vec{p}_h|\cos{\theta}),
\end{equation}
where  $\theta$ is the angle between $\vec{p}_l$ and $\vec{p}_h$.
\end{itemize}
As far as a non-negligible portion of hadrons escapes detection,
 the actual value of $M_h$ is significantly smaller and $M_A$
 correspondingly larger than $M_W$, while $M_{\mathrm{miss}}$ 
 is far from zero.

In a very general case it would be preferable to include all directly
 measured quantities into the analysis, but this is clearly problematic
 because of the huge dimension of the resulting problem. 
 In this analysis we will use just two measured variables: $M_A$ and
\begin{equation}
M_B \equiv \sqrt{M_h^2-M_{\mathrm{miss}}^2}.
\end{equation}
 The better the detector is, the closer the measured
 values of $M_A$ and $M_B$ are to $M_1$ and $M_2$,  correspondingly.  
 Any event from the 2-dimensional plot of the generated masses
 of $W$ bosons $M_1~vs.~M_2$ is transformed by the measurement process to
 an entry into the measured two-dimensional histogram $M_A~vs.~M_B$. To be
 more precise, some of the generated events will be washed away
 because of the limited acceptance.

The overall response matrix of the measurement
 $\hat A_{ij,kl}$ can be defined as the
 probability for an event generated in bin $kl$ of the
 initial $M_1~vs.~M_2$ histogram to find itself in the bin $ij$
 of the measured
 $M_A~vs.~M_B$ histogram. The measurement process is now simulated by
 the multiplication (convolution) of the probability response matrix
 $\hat A_{ij,kl}$ by the generated 2-dimensional histogram
 $H^{(12){\mathrm{sim}}}\equiv M_1~vs.~M_2$,
 yielding the simulated ``measured'' histogram
 $H^{AB,{\mathrm{sim}}}\equiv M_A~vs.~M_B$:
\begin{equation}\label{sim}
\sum_{kl}\hat A_{ij,kl}H^{(12){\mathrm{sim}}}_{kl}=
H^{AB,{\mathrm{sim}}}_{ij}.
\end{equation}

We will assume that the same response matrix relates the true (yet unknown)
two-dimensional mass distribution $H^{(12)}$ to a genuine measured
 histogram $H^{AB}$:
\begin{equation}\label{con}
\sum_{kl}\hat A_{ij,kl}H^{(12)}_{kl}=H^{AB}_{ij}.
\end{equation}
With the known response matrix $\hat{A}$, one can usually fold (convolute) 
 different simulated distributions $H^{(12)}$ with various $M_W$ in order
 to determine which value of $W$ mass corresponds to the measured
 pattern in the r.h.s. Alternatively, one can use the eq.(\ref{con}) to
 unfold $H^{(12)}$ from the measured histogram.
 It is well known that the latter procedure is unstable, and without proper
 stabilization ({\it{regularization}})
 measures would yield totally useless
 rapidly oscillating solutions. However, using the rather simple and
 still very efficient unfolding algorithm developed in {\cite{svd}} it
 is possible to perform the unfolding quite successfully.
 The built-in regularization
 effectively suppresses the spurious oscillations, leaving only
 statistically significant components of the solution.

The unfolding procedure {\cite{svd}} takes the response
 matrix $\hat A$ and the measured r.h.s. $H^{AB}$ as input and 
 outputs the unfolded 2-dimensional mass distribution of the two 
 $W$ bosons $H^{(12)}$, together with its error matrix and the
 inverse of the latter. The unfolded histogram $H^{(12)}$
 can then be projected onto its axes, to obtain separately mass
 distributions for leptonically and hadronically decayed $W$s.      
 As the two distributions must be identical, their sum  
 should be fitted to obtain the $W$ mass. Note that the
 unfolded mass spectrum contains strong bin-to-bin correlations, and
 the inverse of the full error matrix must be used in the fit.

The above analysis can be easily extended to include three or even four
 measured variables  without increasing too much the computing power
 requirements: this extension would increase the number of equations in
 the system ({\ref{con}}) but {\it{not}} the number of unknowns, so that
 the most time-consuming part of the unfolding procedure {\cite{svd}}
 would remain unchanged. 

\section{Example: $W$ mass from single lepton spectrum}\label{exa}

In order to demonstrate the use of the above method
 without involving complicated
 simulations of the detection of hadronic decays,
 a study of a simpler case was performed where the $W$ mass was extracted
 from the shape of the energy spectrum of the single charged lepton.
 Both $W\to e\nu$ and $W\to\mu\nu$ decays have been used. 
 The main sensitivity of the lepton energy spectrum towards 
 the mass of the $W$-boson lies in the
 end-points. In the conventional analysis {\cite{tn3,tn4}}
 the statistical precision even for the most sensitive
 region of energies is not expected to be better than
 300 MeV.

The important aspect of the problem which can be tested in this
 example is the simultaneous dependence of the measured quantity --- the
 charged lepton energy $E_l$ --- upon {\it{both}} $W$ masses in the same event:
\begin{equation}
{{M_1^2}\over{2(E_1-P_1)}}\geq E_l \geq {{M_1^2}\over{2(E_1+P_1)}},
\end{equation}
\begin{equation}
E_1={{s+M_1^2-M_2^2}\over{2\sqrt{s}}},\;\;\;P_1=\sqrt{E_1^2-M_1^2},
\end{equation}
 where $M_i,P_i$ and $E_i$ denote the masses, momenta and energies of the
 $W$-bosons, with the index $i=1$ referring to the parent $W$ and $i=2$ to 
 the recoil one.
 So, an event in the two-dimensional histogram
 $M_1~vs.~M_2$ corresponds to an entry into the
 distribution of the charged lepton energy $E_l$. 
 The convolution equation takes the form:
\begin{equation}\label{col}
\sum_{mn}\hat A^{(l)}_{i,mn}H^{(12)}_{mn}=H^{(l)}_{i},
\end{equation}
with $\hat A^{(l)}$ being the response matrix for each lepton type
$l=e,\mu$, while $H^{(l)}$ is the measured lepton energy distribution.

In order to obtain the matrices $\hat A^{(e)}$ and $\hat A^{(\mu)}$, 
 $2\cdot 10^5$ events of $W$ pair production
 with an electron or a muon in the final state
 have been generated at the energy
 $183$ GeV,  using PYTHIA 5.7 {\cite{tn8}}
 Monte Carlo event generator.
 The detector response was simulated using the following assumptions:
\begin{itemize}
\item{Only leptons within the polar angle $\Theta$ in the
 range $20^0 < \Theta < 160^0$ are detected.}
\item{The energy resolution for electrons is
$\Delta E_{e}/E_{e}=0.2/\sqrt{{E_{e}}(\mathrm{GeV})}$.}
\item{The momentum resolution for muons is
 $\Delta P_{\mu}/P_{\mu}=10^{-3}P_{\mu}(\mathrm{GeV/c})$.}
\end{itemize}

The unfolding method of ref. \cite{svd} requires the introduction of the
 regularization parameter $\tau$, whose optimal value depends on
 both the properties of the response matrix and the statistical accuracy
 of the measured data. The most reliable way of estimating the correct
 $\tau$ in this case is to apply the procedure to a simulated
 spectrum where the true distribution is known, and to minimize the
 deviation of the unfolded distribution from the true one.

Over 40 sets of data at different $W$ mass values from the interval
 $80.25\pm 0.25$ GeV
 have been generated, each consisting of 1700 electrons or muons in the
 final state. This number corresponds to the integrated luminosity of
 500 pb${}^{-1}$ at 183 GeV. The
 generated and the unfolded mass distributions were then compared to
 each other, and
 the optimal values of the regularization parameter
 $\tau$ were obtained, different for $e$ and $\mu$ samples.
 After that, the procedure was applied to independent test sets
 of 1700 events with various $W$ mass values from the same
 interval $80.25 \pm 0.25$ GeV.
 For each sample, the unfolded two-dimensional $W$ mass 
 distribution was projected
 onto the axes and the sum of the projections
 was fitted by a Breit-Wigner parameterization, using 
 the full inverse error matrix.
 An example with $M_W=80.50$ GeV using the muon sample only is 
 presented in Fig.1. 

There are three contributions to the overall error of the $W$ mass
 determined by the above procedure. First,
 the purely statistical error comes from the fit of the unfolded
 distribution:
 $\Delta M_W^{\mathrm{stat}}\approx 250~ {\mathrm{MeV}}$ for the 
 combined $e/\mu$ sample,
 slightly better than the 300 MeV expected from the lepton
 end-point energy measurement.
 Second, the systematic error of the regularized unfolding method connected
 to the choice of the parameter $\tau$:
 $\Delta M_W^{\mathrm{unfo}}\approx100\div 120~ {\mathrm{MeV}}.$
 The last set of uncertainties may come from the inadequate
 event generator program, the inadequate description of the detector and
 the statistical errors in the response matrix itself.
 The first can be estimated by using several Monte Carlo programs,
 while the second requires the detailed understanding of the detector
 response. In this particular example, however, there is no ambiguity in the
 description of the leptonic $W$ decay, and the detector resolution of the
 lepton energy-momentum measurement is expected to be known well enough.
 As for the statistical errors in the response
 matrix, their effects can be eventually made negligible
 compared to the statistical accuracy of the measured data.  In our example
 they do not exceed $10$ MeV.

\section{Discussion}\label{dis}

The example with $W$ mass determination using the single charged lepton
 spectra shows that direct unfolding of the
 immediately measured data into the required distribution is a
 powerful method of data analysis.
 Errors obtained by our procedure
 are slightly smaller compared to those from the end-point
 analysis \cite{tn4} despite the higher c.m.s. energy and accurate accounting
 for the $W$ width in our case --- the two effects which would 
 inevitably make the error of the conventional analysis larger than
 the quoted 300 MeV.
 In the case when the full measured
 information on both the lepton and the hadrons is used, our approach
 can also result in an important
 improvement of accuracy, because of the following reasons:
\begin{itemize}

\item{Each  measured event $e^+e^-\to W^+W^- \to jet+jet+l+\nu$ 
 gives two entries into the final  $W$ mass distribution,
 thus effectively increasing the
 statistics compared to the constrained fit method \cite{tn3,tn4},
 which gives one entry 
 with the ``average mass'' $(M_1+M_2)/2$~
{\footnote{If the overall uncertainty in $M_W$ were dominated by the
gaussian-distributed detector resolution effects the two approaches
would lead to identical statistical errors, as the gaussian
for the average mass would be narrower. However, the significant
portion of the measured $W$ width in the constrained fit method
comes from the natural width of the $W$ boson which obeys the Breit-Wigner
(or indeed Cauchy) distribution. The latter remains invariant under 
the averaging: clearly, the $W$ lifetime should not change as long as 
the two $W$s decay independently.}}
.}
\item{The unfolding restores the natural width
 of the $W$, which is smaller than the expected resolution.}
\item{The use of a well-balanced
 unfolding procedure allows us to retain all statistically significant
 data, simultaneously avoiding the amplification of systematic errors
 which usually takes place if separate
 correction routines are applied at various stages
 of the analysis.}

\end{itemize}

The most important systematic error specific to the direct
 unfolding method itself comes from the uncertainty in the determination
 of the regularization parameter $\tau$. This error can be reliably
 estimated using various simulated ``measured'' distributions, and can
 be shown to decrease when the statistical significance of the measured
 data increases \cite{svd}. The effects due to the initial state
 radiation can be either included into the response matrix (as we have
 done in our example), or accommodated at a later stage of the
 analysis. A more subtle source of errors is the
 response matrix: any deviation of the Monte Carlo event generator
 or the detector model from reality may potentially result in a
 systematic effect in the unfolded distribution. However, we
 believe that the huge experimental data accumulated by the four LEP
 collaborations will eventually
 allow one to minimize those uncertainties. The hadronic jets in
 $W$ decays should be quite similar to the ones originating from
 the $Z$ peak, and the variation in the jet energy can be
 accounted for by the proper Lorentz boost. Finally, the effects
 of the statistical fluctuations in the response matrix can be
 made negligible compared to the influence of the
 statistical errors in the measured
 data by the proper increase of the generated statistics.
 In any case, by combining the results
 of various methods of analysis one can achieve 
 better understanding of the overall
 uncertainty in $W$ mass determination.

\section{Acknowlegdements}
 We are grateful to R. Barlow, N. Kjaer, A. Lebedev, A. Olshevski, T. Shears
 and  D. Ward for helpful discussions. We would like to thank DELPHI
 collaboration and CERN for their kind hospitality during our stay
 at CERN, where this work has been finalized.

\newpage
                                                       
\begin{figure}[top]
\begin{center}
\mbox{
      \epsfysize=16cm
      \epsfbox{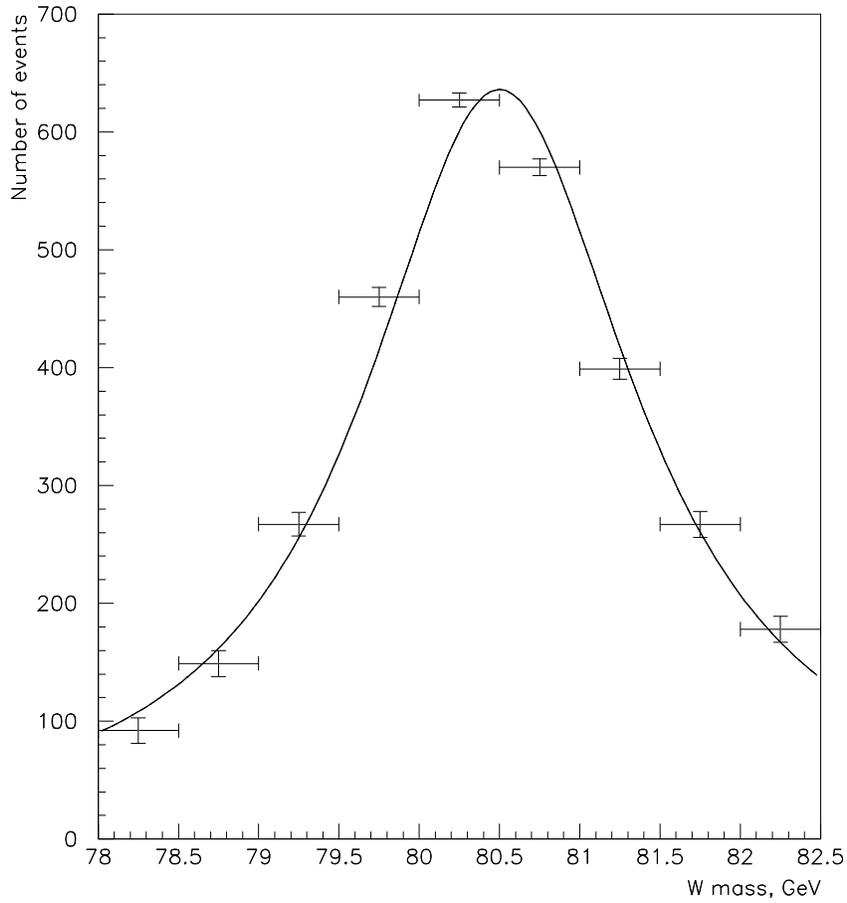}
      }
\end{center}            

\caption[1]{$W$ mass distribution unfolded from a sample
of 1700 simulated events with final state muons.
The errors shown correspond to the diagonal elements of the error matrix.
The smooth curve describes the fit performed using 
the full error matrix including bin-to-bin
correlations, resulting in $M_W=80.50\pm 0.33$ GeV.}

\end{figure}            

\end{document}